
%
\expandafter\ifx\csname phyzzx\endcsname\relax
 \message{It is better to use PHYZZX format than to
          \string\input\space PHYZZX}\else
 \wlog{PHYZZX macros are already loaded and are not
          \string\input\space again}%
 \endinput \fi
\catcode`\@=11 
\let\rel@x=\relax
\let\n@expand=\relax
\def\pr@tect{\let\n@expand=\noexpand}
\let\protect=\pr@tect
\let\gl@bal=\global
%
%
%
\newfam\cpfam
\newdimen\b@gheight             \b@gheight=12pt
\newcount\f@ntkey               \f@ntkey=0
\def\f@m{\afterassignment\samef@nt\f@ntkey=}
\def\samef@nt{\fam=\f@ntkey \the\textfont\f@ntkey\rel@x}
\def\setstr@t{\setbox\strutbox=\hbox{\vrule height 0.85\b@gheight
                                depth 0.35\b@gheight width\z@ }}
%
%
%
%
%

\font\fourteenrm  =cmr10 scaled\magstep2
\font\twelverm    =cmr12
\font\ninerm      =cmr9
\font\sixrm       =cmr6

\font\fourteenbf  =cmbx10 scaled\magstep2
\font\twelvebf    =cmbx12
\font\ninebf      =cmbx9
\font\sixbf       =cmbx6
\font\seventeeni  =cmmi10 scaled\magstep3    \skewchar\seventeeni='177
\font\fourteeni   =cmmi10 scaled\magstep2     \skewchar\fourteeni='177
\font\twelvei     =cmmi12                       \skewchar\twelvei='177
\font\ninei       =cmmi9                          \skewchar\ninei='177
\font\sixi        =cmmi6                           \skewchar\sixi='177
\font\seventeensy =cmsy10 scaled\magstep3    \skewchar\seventeensy='60
\font\fourteensy  =cmsy10 scaled\magstep2     \skewchar\fourteensy='60
\font\twelvesy    =cmsy10 scaled\magstep1       \skewchar\twelvesy='60
\font\ninesy      =cmsy9                          \skewchar\ninesy='60
\font\sixsy       =cmsy6                           \skewchar\sixsy='60

\font\fourteenex  =cmex10 scaled\magstep2
\font\twelveex    =cmex10 scaled\magstep1

\font\fourteensl  =cmsl10 scaled\magstep2
\font\twelvesl    =cmsl12
\font\ninesl      =cmsl9

\font\fourteenit  =cmti10 scaled\magstep2
\font\twelveit    =cmti12
\font\nineit      =cmti9
\font\fourteentt  =cmtt10 scaled\magstep2
\font\twelvett    =cmtt12
\font\fourteencp  =cmcsc10 scaled\magstep2
\font\twelvecp    =cmcsc10 scaled\magstep1
\font\tencp       =cmcsc10
%
%
\def\fourteenf@nts{\relax
    \textfont0=\fourteenrm          \scriptfont0=\tenrm
      \scriptscriptfont0=\sevenrm
    \textfont1=\fourteeni           \scriptfont1=\teni
      \scriptscriptfont1=\seveni
    \textfont2=\fourteensy          \scriptfont2=\tensy
      \scriptscriptfont2=\sevensy
    \textfont3=\fourteenex          \scriptfont3=\twelveex
      \scriptscriptfont3=\tenex
    \textfont\itfam=\fourteenit     \scriptfont\itfam=\tenit
    \textfont\slfam=\fourteensl     \scriptfont\slfam=\tensl
    \textfont\bffam=\fourteenbf     \scriptfont\bffam=\tenbf
      \scriptscriptfont\bffam=\sevenbf
    \textfont\ttfam=\fourteentt
    \textfont\cpfam=\fourteencp }
\def\twelvef@nts{\relax
    \textfont0=\twelverm          \scriptfont0=\ninerm
      \scriptscriptfont0=\sixrm
    \textfont1=\twelvei           \scriptfont1=\ninei
      \scriptscriptfont1=\sixi
    \textfont2=\twelvesy           \scriptfont2=\ninesy
      \scriptscriptfont2=\sixsy
    \textfont3=\twelveex          \scriptfont3=\tenex
      \scriptscriptfont3=\tenex
    \textfont\itfam=\twelveit     \scriptfont\itfam=\nineit
    \textfont\slfam=\twelvesl     \scriptfont\slfam=\ninesl
    \textfont\bffam=\twelvebf     \scriptfont\bffam=\ninebf
      \scriptscriptfont\bffam=\sixbf
    \textfont\ttfam=\twelvett
    \textfont\cpfam=\twelvecp }
\def\tenf@nts{\relax
    \textfont0=\tenrm          \scriptfont0=\sevenrm
      \scriptscriptfont0=\fiverm
    \textfont1=\teni           \scriptfont1=\seveni
      \scriptscriptfont1=\fivei
    \textfont2=\tensy          \scriptfont2=\sevensy
      \scriptscriptfont2=\fivesy
    \textfont3=\tenex          \scriptfont3=\tenex
      \scriptscriptfont3=\tenex
    \textfont\itfam=\tenit     \scriptfont\itfam=\seveni  
    \textfont\slfam=\tensl     \scriptfont\slfam=\sevenrm 
    \textfont\bffam=\tenbf     \scriptfont\bffam=\sevenbf
      \scriptscriptfont\bffam=\fivebf
    \textfont\ttfam=\tentt
    \textfont\cpfam=\tencp }
%
%

%
\def\rm{\n@expand\f@m0 }
\def\mit{\n@expand\f@m1 }         
\def\cal{\n@expand\f@m2 }
\def\it{\n@expand\f@m\itfam}
\def\sl{\n@expand\f@m\slfam}
\def\bf{\n@expand\f@m\bffam}
\def\tt{\n@expand\f@m\ttfam}
\def\caps{\n@expand\f@m\cpfam}    
\def\em@{\rel@x\ifnum\f@ntkey=0 \it \else
        \ifnum\f@ntkey=\bffam \it \else \rm \fi \fi }
\def\em{\n@expand\em@}
\def\fourteenpoint{\fourteenf@nts \samef@nt \b@gheight=14pt \setstr@t }
\def\twelvepoint{\twelvef@nts \samef@nt \b@gheight=12pt \setstr@t }
\def\tenpoint{\tenf@nts \samef@nt \b@gheight=10pt \setstr@t }
\normalbaselineskip = 19.2pt plus 0.2pt minus 0.1pt 
\normallineskip = 1.5pt plus 0.1pt minus 0.1pt
\normallineskiplimit = 1.5pt
\newskip\normaldisplayskip
\normaldisplayskip = 14.4pt plus 3.6pt minus 10.0pt 
\newskip\normaldispshortskip
\normaldispshortskip = 6pt plus 5pt
\newskip\normalparskip
\normalparskip = 6pt plus 2pt minus 1pt
\newskip\skipregister
\skipregister = 5pt plus 2pt minus 1.5pt
\newif\ifsingl@
\newif\ifdoubl@
\newif\iftwelv@  \twelv@true
\def\singlespace{\singl@true\doubl@false\spaces@t}
\def\doublespace{\singl@false\doubl@true\spaces@t}
\def\normalspace{\singl@false\doubl@false\spaces@t}
\def\Tenpoint{\tenpoint\twelv@false\spaces@t}
\def\Twelvepoint{\twelvepoint\twelv@true\spaces@t}
\def\spaces@t{\rel@x
      \iftwelv@ \ifsingl@\subspaces@t3:4;\else\subspaces@t1:1;\fi
       \else \ifsingl@\subspaces@t3:5;\else\subspaces@t4:5;\fi \fi
      \ifdoubl@ \multiply\baselineskip by 5
         \divide\baselineskip by 4 \fi }
\def\subspaces@t#1:#2;{
      \baselineskip = \normalbaselineskip
      \multiply\baselineskip by #1 \divide\baselineskip by #2
      \lineskip = \normallineskip
      \multiply\lineskip by #1 \divide\lineskip by #2
      \lineskiplimit = \normallineskiplimit
      \multiply\lineskiplimit by #1 \divide\lineskiplimit by #2
      \parskip = \normalparskip
      \multiply\parskip by #1 \divide\parskip by #2
      \abovedisplayskip = \normaldisplayskip
      \multiply\abovedisplayskip by #1 \divide\abovedisplayskip by #2
      \belowdisplayskip = \abovedisplayskip
      \abovedisplayshortskip = \normaldispshortskip
      \multiply\abovedisplayshortskip by #1
        \divide\abovedisplayshortskip by #2
      \belowdisplayshortskip = \abovedisplayshortskip
      \advance\belowdisplayshortskip by \belowdisplayskip
      \divide\belowdisplayshortskip by 2
      \smallskipamount = \skipregister
      \multiply\smallskipamount by #1 \divide\smallskipamount by #2
      \medskipamount = \smallskipamount \multiply\medskipamount by 2
      \bigskipamount = \smallskipamount \multiply\bigskipamount by 4 }
\def\normalbaselines{ \baselineskip=\normalbaselineskip
   \lineskip=\normallineskip \lineskiplimit=\normallineskip
   \iftwelv@\else \multiply\baselineskip by 4 \divide\baselineskip by 5
     \multiply\lineskiplimit by 4 \divide\lineskiplimit by 5
     \multiply\lineskip by 4 \divide\lineskip by 5 \fi }
\Twelvepoint  
\interlinepenalty=50
\interfootnotelinepenalty=5000
\predisplaypenalty=9000
\postdisplaypenalty=500
\hfuzz=1pt
\vfuzz=0.2pt
\newdimen\HOFFSET  \HOFFSET=0pt
\newdimen\VOFFSET  \VOFFSET=0pt
\newdimen\HSWING   \HSWING=0pt
\dimen\footins=8in
%
%
%
\newskip\pagebottomfiller
\pagebottomfiller=\z@ plus \z@ minus \z@
\def\pagecontents{
   \ifvoid\topins\else\unvbox\topins\vskip\skip\topins\fi
   \dimen@ = \dp255 \unvbox255
   \vskip\pagebottomfiller
   \ifvoid\footins\else\vskip\skip\footins\footrule\unvbox\footins\fi
   \ifr@ggedbottom \kern-\dimen@ \vfil \fi }
\def\makeheadline{\vbox to 0pt{ \skip@=\topskip
      \advance\skip@ by -12pt \advance\skip@ by -2\normalbaselineskip
      \vskip\skip@ \line{\vbox to 12pt{}\the\headline} \vss
      }\nointerlineskip}
\def\makefootline{\baselineskip = 1.5\normalbaselineskip
                 \line{\the\footline}}
\newif\iffrontpage
\newif\ifp@genum
\def\nopagenumbers{\p@genumfalse}
\def\pagenumbers{\p@genumtrue}
\pagenumbers
\newtoks\paperheadline
\newtoks\paperfootline
\newtoks\letterheadline
\newtoks\letterfootline
\newtoks\letterinfo
\newtoks\date
\paperheadline={\hfil}
\paperfootline={\hss\iffrontpage\else\ifp@genum\tenrm\folio\hss\fi\fi}
\letterheadline{\iffrontpage \hfil \else
    \rm \ifp@genum page~~\folio\fi \hfil\the\date \fi}
\letterfootline={\iffrontpage\the\letterinfo\else\hfil\fi}
\letterinfo={\hfil}
\def\monthname{\rel@x\ifcase\month 0/\or January\or February\or
   March\or April\or May\or June\or July\or August\or September\or
   October\or November\or December\else\number\month/\fi}

\date={}
\headline=\paperheadline 
\footline=\paperfootline 
\countdef\pageno=1      \countdef\pagen@=0
\countdef\pagenumber=1  \pagenumber=1
\def\advancepageno{\gl@bal\advance\pagen@ by 1
   \ifnum\pagenumber<0 \gl@bal\advance\pagenumber by -1
    \else\gl@bal\advance\pagenumber by 1 \fi
    \gl@bal\frontpagefalse  \swing@ }
\def\folio{\ifnum\pagenumber<0 \romannumeral-\pagenumber
           \else \number\pagenumber \fi }
\def\swing@{\ifodd\pagenumber \gl@bal\advance\hoffset by -\HSWING
             \else \gl@bal\advance\hoffset by \HSWING \fi }
\def\footrule{\dimen@=\prevdepth\nointerlineskip
   \vbox to 0pt{\vskip -0.25\baselineskip \hrule width 0.35\hsize \vss}
   \prevdepth=\dimen@ }
\let\footnotespecial=\rel@x
\newdimen\footindent
\footindent=24pt
\def\Textindent#1{\noindent\llap{#1\enspace}\ignorespaces}
\def\Vfootnote#1{\insert\footins\bgroup
   \interlinepenalty=\interfootnotelinepenalty \floatingpenalty=20000
   \singl@true\doubl@false\Tenpoint
   \splittopskip=\ht\strutbox \boxmaxdepth=\dp\strutbox
   \leftskip=\footindent \rightskip=\z@skip
   \parindent=0.5\footindent \parfillskip=0pt plus 1fil
   \spaceskip=\z@skip \xspaceskip=\z@skip \footnotespecial
   \Textindent{#1}\footstrut\futurelet\next\fo@t}

\def\vfootnote#1{\Vfootnote{${#1}$}}
\def\footnote#1{\attach{#1}\vfootnote{#1}}

\def\foot{\attach\footsymbolgen\vfootnote{\footsymbol}}
\let\footsymbol=\star
\newcount\lastf@@t           \lastf@@t=-1
\newcount\footsymbolcount    \footsymbolcount=0
\newif\ifPhysRev
\def\footsymbolgen{\bumpfootsymbolcount \generatefootsymbol \footsymbol }
\def\bumpfootsymbolcount{\rel@x
   \iffrontpage \bumpfootsymbolpos \else \advance\lastf@@t by 1
     \ifPhysRev \bumpfootsymbolneg \else \bumpfootsymbolpos \fi \fi
   \gl@bal\lastf@@t=\pagen@ }
\def\bumpfootsymbolpos{\ifnum\footsymbolcount <0
                            \gl@bal\footsymbolcount =0 \fi
    \ifnum\lastf@@t<\pagen@ \gl@bal\footsymbolcount=0
     \else \gl@bal\advance\footsymbolcount by 1 \fi }
\def\bumpfootsymbolneg{\ifnum\footsymbolcount >0
             \gl@bal\footsymbolcount =0 \fi
         \gl@bal\advance\footsymbolcount by -1 }
\def\fd@f#1 {\xdef\footsymbol{\mathchar"#1 }}
\def\generatefootsymbol{\ifcase\footsymbolcount \fd@f 13F \or \fd@f 279
        \or \fd@f 27A \or \fd@f 278 \or \fd@f 27B \else
        \ifnum\footsymbolcount <0 \fd@f{023 \number-\footsymbolcount }
         \else \fd@f 203 {\loop \ifnum\footsymbolcount >5
                \fd@f{203 \footsymbol } \advance\footsymbolcount by -1
                \repeat }\fi \fi }

\def\nonfrenchspacing{\sfcode`\.=3001 \sfcode`\!=3000 \sfcode`\?=3000
        \sfcode`\:=2000 \sfcode`\;=1500 \sfcode`\,=1251 }
\nonfrenchspacing
\newdimen\d@twidth
{\setbox0=\hbox{s.} \gl@bal\d@twidth=\wd0 \setbox0=\hbox{s}
        \gl@bal\advance\d@twidth by -\wd0 }
\def\removehglue{\loop \unskip \ifdim\lastskip >\z@ \repeat }
\def\roll@ver#1{\removehglue \nobreak \count255 =\spacefactor \dimen@=\z@
        \ifnum\count255 =3001 \dimen@=\d@twidth \fi
        \ifnum\count255 =1251 \dimen@=\d@twidth \fi
    \iftwelv@ \kern-\dimen@ \else \kern-0.83\dimen@ \fi
   #1\spacefactor=\count255 }
\def\step@ver#1{\rel@x \ifmmode #1\else \ifhmode
        \roll@ver{${}#1$}\else {\setbox0=\hbox{${}#1$}}\fi\fi }
\def\attach#1{\step@ver{\strut^{\mkern 2mu #1} }}
%
%
%
\newcount\chapternumber      \chapternumber=0
\newcount\sectionnumber      \sectionnumber=0
\newcount\equanumber         \equanumber=0
\let\chapterlabel=\rel@x
\let\sectionlabel=\rel@x
\newtoks\chapterstyle        \chapterstyle={\Number}
\newtoks\sectionstyle        \sectionstyle={\chapterlabel.\Number}
\newskip\chapterskip         \chapterskip=\bigskipamount
\newskip\sectionskip         \sectionskip=\medskipamount
\newskip\headskip            \headskip=8pt plus 3pt minus 3pt
\newdimen\chapterminspace    \chapterminspace=15pc
\newdimen\sectionminspace    \sectionminspace=10pc
\newdimen\referenceminspace  \referenceminspace=20pc
\def\chapterreset{\gl@bal\advance\chapternumber by 1
   \ifnum\equanumber<0 \else\gl@bal\equanumber=0\fi
   \sectionnumber=0 \let\sectionlabel=\rel@x
   {\pr@tect\xdef\chapterlabel{\the\chapterstyle{\the\chapternumber}}}}
\def\alphabetic#1{\count255='140 \advance\count255 by #1\char\count255}
\def\Alphabetic#1{\count255='100 \advance\count255 by #1\char\count255}
\def\Roman#1{\uppercase\expandafter{\romannumeral #1}}
\def\roman#1{\romannumeral #1}
\def\Number#1{\number #1}
\def\BLANC#1{}
\def\titleparagraphs{\interlinepenalty=9999
     \leftskip=0.03\hsize plus 0.22\hsize minus 0.03\hsize
     \rightskip=\leftskip \parfillskip=0pt
     \hyphenpenalty=9000 \exhyphenpenalty=9000
     \tolerance=9999 \pretolerance=9000
     \spaceskip=0.333em \xspaceskip=0.5em }
\def\titlestyle#1{\par\begingroup \titleparagraphs
     \iftwelv@\fourteenpoint\else\twelvepoint\fi
   \noindent #1\par\endgroup }
\def\spacecheck#1{\dimen@=\pagegoal\advance\dimen@ by -\pagetotal
   \ifdim\dimen@<#1 \ifdim\dimen@>0pt \vfil\break \fi\fi}
\def\chapter#1{\par \penalty-300 \vskip\chapterskip
   \spacecheck\chapterminspace
   \chapterreset \titlestyle{\chapterlabel.~#1}
   \nobreak\vskip\headskip \penalty 30000
   {\pr@tect\wlog{\string\chapter\space \chapterlabel}} }

\def\section#1{\par \ifnum\the\lastpenalty=30000\else
   \penalty-200\vskip\sectionskip \spacecheck\sectionminspace\fi
   \gl@bal\advance\sectionnumber by 1
   {\pr@tect
   \xdef\sectionlabel{\the\sectionstyle\the\sectionnumber}
   \wlog{\string\section\space \sectionlabel}}
   \noindent {\caps\enspace\sectionlabel.~~#1}\par
   \nobreak\vskip\headskip \penalty 30000 }
\def\subsection#1{\par
   \ifnum\the\lastpenalty=30000\else \penalty-100\smallskip \fi
   \noindent\undertext{#1}\enspace \vadjust{\penalty5000}}

\def\undertext#1{\vtop{\hbox{#1}\kern 1pt \hrule}}
\def\APPENDIX#1#2{\par\penalty-300\vskip\chapterskip
   \spacecheck\chapterminspace \chapterreset \xdef\chapterlabel{#1}
   \titlestyle{APPENDIX #2} \nobreak\vskip\headskip \penalty 30000
   \wlog{\string\Appendix~\chapterlabel} }
\def\Appendix#1{\APPENDIX{#1}{#1}}
\def\appendix{\APPENDIX{A}{}}
\def\unnumberedchapters{\let\makechapterlabel=\rel@x
      \let\chapterlabel=\rel@x  \sectionstyle={\BLANC}
      \let\sectionlabel=\rel@x \sequentialequations }
%
%
%
\def\eqname#1{\rel@x {\pr@tect
  \ifnum\equanumber<0 \xdef#1{{\rm(\number-\equanumber)}}%
     \gl@bal\advance\equanumber by -1
  \else \gl@bal\advance\equanumber by 1
     \ifx\chapterlabel\rel@x \def\d@t{}\else \def\d@t{.}\fi
    \xdef#1{{\rm(\chapterlabel\d@t\number\equanumber)}}\fi #1}}

\def\eqn{\eqno\eqname}

\def\eqinsert#1{\noalign{\dimen@=\prevdepth \nointerlineskip
   \setbox0=\hbox to\displaywidth{\hfil #1}
   \vbox to 0pt{\kern 0.5\baselineskip\hbox{$\!\box0\!$}\vss}
   \prevdepth=\dimen@}}
%

%
%
\def\GENITEM#1;#2{\par \hangafter=0 \hangindent=#1
    \Textindent{$ #2 $}\ignorespaces}
\outer\def\newitem#1=#2;{\gdef#1{\GENITEM #2;}}

\newdimen\itemsize                \itemsize=30pt
\newitem\item=1\itemsize;
\newitem\sitem=1.75\itemsize;     
\newitem\ssitem=2.5\itemsize;     
\outer\def\newlist#1=#2&#3&#4;{\toks0={#2}\toks1={#3}%
   \count255=\escapechar \escapechar=-1
   \alloc@0\list\countdef\insc@unt\listcount     \listcount=0
   \edef#1{\par
      \countdef\listcount=\the\allocationnumber
      \advance\listcount by 1
      \hangafter=0 \hangindent=#4
      \Textindent{\the\toks0{\listcount}\the\toks1}}
   \expandafter\expandafter\expandafter
    \edef\c@t#1{begin}{\par
      \countdef\listcount=\the\allocationnumber \listcount=1
      \hangafter=0 \hangindent=#4
      \Textindent{\the\toks0{\listcount}\the\toks1}}
   \expandafter\expandafter\expandafter
    \edef\c@t#1{con}{\par \hangafter=0 \hangindent=#4 \noindent}
   \escapechar=\count255}
\def\c@t#1#2{\csname\string#1#2\endcsname}
\newlist\point=\Number&.&1.0\itemsize;
\newlist\subpoint=(\alphabetic&)&1.75\itemsize;
\newlist\subsubpoint=(\roman&)&2.5\itemsize;
%

%
%
%
%
\newcount\referencecount     \referencecount=0
\newcount\lastrefsbegincount \lastrefsbegincount=0
\newif\ifreferenceopen       \newwrite\referencewrite
\newdimen\refindent          \refindent=30pt
\def\normalrefmark#1{\attach{\scriptscriptstyle [ #1 ] }}
\let\PRrefmark=\attach
\def\NPrefmark#1{\step@ver{{\;[#1]}}}
\def\refmark#1{\rel@x\ifPhysRev\PRrefmark{#1}\else\normalrefmark{#1}\fi}
\def\refend@{\refmark{\number\referencecount}}
\def\refend{\refend@{}\space }
\def\refsend{\refmark{\count255=\referencecount
   \advance\count255 by-\lastrefsbegincount
   \ifcase\count255 \number\referencecount
   \or \number\lastrefsbegincount,\number\referencecount
   \else \number\lastrefsbegincount-\number\referencecount \fi}\space }
\def\REFNUM#1{\rel@x \gl@bal\advance\referencecount by 1
    \xdef#1{\the\referencecount }}
\def\Refnum#1{\REFNUM #1\refend@ } 
\def\REF#1{\REFNUM #1\R@FWRITE\ignorespaces}
\def\Ref#1{\Refnum #1\REFWRITE }
\def\ref{\Ref\?}
\def\REFS#1{\REFNUM #1\gl@bal\lastrefsbegincount=\referencecount
    \REFWRITE }

\def\r@fitem#1{\par \hangafter=0 \hangindent=\refindent \Textindent{#1}}
\def\refitem#1{\r@fitem{#1.}}
\def\NPrefitem#1{\r@fitem{[#1]}}
\def\NPrefs{\let\refmark=\NPrefmark \let\refitem=\NPrefitem}
\def\REFWRITE{\R@FWRITE\rel@x }
\def\R@FWRITE#1{\ifreferenceopen \else \gl@bal\referenceopentrue
     \immediate\openout\referencewrite=\jobname.refs
     \toks@={\begingroup \refoutspecials \catcode`\^^M=10 }%
     \immediate\write\referencewrite{\the\toks@}\fi
    \immediate\write\referencewrite{\noexpand\refitem %
                                    {\the\referencecount}}%
    \p@rse@ndwrite \referencewrite #1}
\begingroup
 \catcode`\^^M=\active \let^^M=\relax %
 \gdef\p@rse@ndwrite#1#2{\begingroup \catcode`\^^M=12 \newlinechar=`\^^M%
         \chardef\rw@write=#1\sc@nlines#2}%
 \gdef\sc@nlines#1#2{\sc@n@line \g@rbage #2^^M\endsc@n \endgroup #1}%
 \gdef\sc@n@line#1^^M{\expandafter\toks@\expandafter{\deg@rbage #1}%
         \immediate\write\rw@write{\the\toks@}%
         \futurelet\n@xt \sc@ntest }%
\endgroup
\def\sc@ntest{\ifx\n@xt\endsc@n \let\n@xt=\rel@x
       \else \let\n@xt=\sc@n@notherline \fi \n@xt }
\def\sc@n@notherline{\sc@n@line \g@rbage }
\def\deg@rbage#1{}
\let\g@rbage=\relax    \let\endsc@n=\relax
\def\refout{\par\penalty-400\vskip\chapterskip
   \spacecheck\referenceminspace
   \ifreferenceopen \Closeout\referencewrite \referenceopenfalse \fi
   \line{\fourteenrm\hfil REFERENCES\hfil}\vskip\headskip
   \input \jobname.refs
   }
\def\refoutspecials{\sfcode`\.=1000 \interlinepenalty=1000
         \rightskip=\z@ plus 1em minus \z@ }
\def\Closeout#1{\toks0={\par\endgroup}\immediate\write#1{\the\toks0}%
   \immediate\closeout#1}
%
%
\newcount\figurecount     \figurecount=0
\newcount\tablecount      \tablecount=0
\newif\iffigureopen       \newwrite\figurewrite
\newif\iftableopen        \newwrite\tablewrite
\def\FIGNUM#1{\rel@x \gl@bal\advance\figurecount by 1
    \xdef#1{\the\figurecount}}
\def\FIGURE#1{\FIGNUM #1\F@GWRITE\ignorespaces }

\def\figitem#1{\r@fitem{#1)}}
\def\FIGWRITE{\F@GWRITE\rel@x }
\def\TABNUM#1{\rel@x \gl@bal\advance\tablecount by 1
    \xdef#1{\the\tablecount}}
\def\TABLE#1{\TABNUM #1\T@BWRITE\ignorespaces }

\def\tabitem#1{\r@fitem{#1:}}
\def\TABWRITE{\T@BWRITE\rel@x }
\def\F@GWRITE#1{\iffigureopen \else \gl@bal\figureopentrue
     \immediate\openout\figurewrite=\jobname.figs
     \toks@={\begingroup \catcode`\^^M=10 }%
     \immediate\write\figurewrite{\the\toks@}\fi
    \immediate\write\figurewrite{\noexpand\figitem %
                                 {\the\figurecount}}%
    \p@rse@ndwrite \figurewrite #1}
\def\T@BWRITE#1{\iftableopen \else \gl@bal\tableopentrue
     \immediate\openout\tablewrite=\jobname.tabs
     \toks@={\begingroup \catcode`\^^M=10 }%
     \immediate\write\tablewrite{\the\toks@}\fi
    \immediate\write\tablewrite{\noexpand\tabitem %
                                 {\the\tablecount}}%
    \p@rse@ndwrite \tablewrite #1}
\def\figout{\par\penalty-400
   \vskip\chapterskip\spacecheck\referenceminspace
   \iffigureopen \Closeout\figurewrite \figureopenfalse \fi
   \line{\fourteenrm\hfil FIGURE CAPTIONS\hfil}\vskip\headskip
   \input \jobname.figs
   }
\def\tabout{\par\penalty-400
   \vskip\chapterskip\spacecheck\referenceminspace
   \iftableopen \Closeout\tablewrite \tableopenfalse \fi
   \line{\fourteenrm\hfil TABLE CAPTIONS\hfil}\vskip\headskip
   \input \jobname.tabs
   }
%
%
%
\newbox\picturebox
\def\p@cht{\ht\picturebox }
\def\p@cwd{\wd\picturebox }
\def\p@cdp{\dp\picturebox }
\newdimen\xshift
\newdimen\yshift
\newdimen\captionwidth
\newskip\captionskip
\captionskip=15pt plus 5pt minus 3pt
\def\fullwidth{\captionwidth=\hsize }
\newtoks\Caption
\newif\ifcaptioned
\newif\ifselfcaptioned
\def\caption{\captionedtrue \Caption }
\newcount\linesabove
\newif\iffileexists
\newtoks\picfilename
\def\fil@#1 {\fileexiststrue \picfilename={#1}}
\def\file#1{\if=#1\let\n@xt=\fil@ \else \def\n@xt{\fil@ #1}\fi \n@xt }
\def\pl@t{\begingroup \pr@tect
    \setbox\picturebox=\hbox{}\fileexistsfalse
    \let\height=\p@cht \let\width=\p@cwd \let\depth=\p@cdp
    \xshift=\z@ \yshift=\z@ \captionwidth=\z@
    \Caption={}\captionedfalse
    \linesabove =0 \picturedefault }
\def\plot{\pl@t \selfcaptionedfalse }
\def\Picture#1{\gl@bal\advance\figurecount by 1
    \xdef#1{\the\figurecount}\pl@t \selfcaptionedtrue }

\def\s@vepicture{\iffileexists \parsefilename \redopicturebox \fi
   \ifdim\captionwidth>\z@ \else \captionwidth=\p@cwd \fi
   \xdef\lastpicture{\iffileexists
        \setbox0=\hbox{\raise\the\yshift \vbox{%
              \moveright\the\xshift\hbox{\picturedefinition}}}%
        \else \setbox0=\hbox{}\fi
         \ht0=\the\p@cht \wd0=\the\p@cwd \dp0=\the\p@cdp
         \vbox{\hsize=\the\captionwidth \line{\hss\box0 \hss }%
              \ifcaptioned \vskip\the\captionskip \noexpand\Tenpoint
                \ifselfcaptioned Figure~\the\figurecount.\enspace \fi
                \the\Caption \fi }}%
    \endgroup }
\let\endpicture=\s@vepicture
\def\savepicture#1{\s@vepicture \global\let#1=\lastpicture }
\def\displaypicture{\fullwidth \s@vepicture $$\lastpicture $${}}
\def\toppicture{\fullwidth \s@vepicture \topinsert
    \lastpicture \medskip \endinsert }
\def\midpicture{\fullwidth \s@vepicture \midinsert
    \lastpicture \endinsert }
%
%
\def\leftpicture{\pres@tpicture
    \dimen@i=\hsize \advance\dimen@i by -\dimen@ii
    \setbox\picturebox=\hbox to \hsize {\box0 \hss }%
    \wr@paround }
\def\rightpicture{\pres@tpicture
    \dimen@i=\z@
    \setbox\picturebox=\hbox to \hsize {\hss \box0 }%
    \wr@paround }
\def\pres@tpicture{\gl@bal\linesabove=\linesabove
    \s@vepicture \setbox\picturebox=\vbox{
         \kern \linesabove\baselineskip \kern 0.3\baselineskip
         \lastpicture \kern 0.3\baselineskip }%
    \dimen@=\p@cht \dimen@i=\dimen@
    \advance\dimen@i by \pagetotal
    \par \ifdim\dimen@i>\pagegoal \vfil\break \fi
    \dimen@ii=\hsize
    \advance\dimen@ii by -\parindent \advance\dimen@ii by -\p@cwd
    \setbox0=\vbox to\z@{\kern-\baselineskip \unvbox\picturebox \vss }}
\def\wr@paround{\Caption={}\count255=1
    \loop \ifnum \linesabove >0
         \advance\linesabove by -1 \advance\count255 by 1
         \advance\dimen@ by -\baselineskip
         \expandafter\Caption \expandafter{\the\Caption \z@ \hsize }%
      \repeat
    \loop \ifdim \dimen@ >\z@
         \advance\count255 by 1 \advance\dimen@ by -\baselineskip
         \expandafter\Caption \expandafter{%
             \the\Caption \dimen@i \dimen@ii }%
      \repeat
    \edef\n@xt{\parshape=\the\count255 \the\Caption \z@ \hsize }%
    \par\noindent \n@xt \strut \vadjust{\box\picturebox }}
\let\picturedefault=\relax
\let\parsefilename=\relax
\def\redopicturebox{\let\picturedefinition=\rel@x
   \errhelp=\disabledpictures
   \errmessage{This version of TeX cannot handle pictures.  Sorry.}}
\newhelp\disabledpictures
     {You will get a blank box in place of your picture.}
%
%
%
%
%
%
%
%
%
%
\def\FRONTPAGE{\ifvoid255\else\vfill\penalty-20000\fi
   \gl@bal\pagenumber=1     \gl@bal\chapternumber=0
   \gl@bal\equanumber=0     \gl@bal\sectionnumber=0
   \gl@bal\referencecount=0 \gl@bal\figurecount=0
   \gl@bal\tablecount=0     \gl@bal\frontpagetrue
   \gl@bal\lastf@@t=0       \gl@bal\footsymbolcount=0}

\def\papers{\papersize\headline=\paperheadline\footline=\paperfootline}
\def\papersize{
   \advance\hoffset by\HOFFSET \advance\voffset by\VOFFSET
   \pagebottomfiller=0pc
   \skip\footins=\bigskipamount \normalspace }
\papers  
%
%
\newskip\lettertopskip       \lettertopskip=20pt plus 50pt
\newskip\letterbottomskip    \letterbottomskip=\z@ plus 100pt
\newskip\signatureskip       \signatureskip=40pt plus 3pt
\def\lettersize{\hsize=6.5in \vsize=8.5in \hoffset=0in \voffset=0.5in
   \advance\hoffset by\HOFFSET \advance\voffset by\VOFFSET
   \pagebottomfiller=\letterbottomskip
   \skip\footins=\smallskipamount \multiply\skip\footins by 3
   \singlespace }
\def\MEMO{\lettersize \headline=\letterheadline \footline={\hfil }%
   \let\rule=\memorule \FRONTPAGE \memohead }

\def\memodate{\afterassignment\MEMO \date }
\def\memit@m#1{\smallskip \hangafter=0 \hangindent=1in
    \Textindent{\caps #1}}
\def\subject{\memit@m{Subject:}}
\def\topic{\memit@m{Topic:}}
\def\from{\memit@m{From:}}
\def\memorule{\medskip\hrule height 1pt\bigskip}  
\def\memohead{\centerline{\fourteenrm MEMORANDUM}}
\newwrite\labelswrite
\newtoks\rw@toks
\def\letters{\lettersize
   \headline=\letterheadline \footline=\letterfootline
   \immediate\openout\labelswrite=\jobname.lab}

\let\letterhead=\rel@x
\def\addressee#1{\medskip\line{\hskip 0.75\hsize plus\z@ minus 0.25\hsize
                               \the\date \hfil }%
   \vskip \lettertopskip
   \ialign to\hsize{\strut ##\hfil\tabskip 0pt plus \hsize \crcr #1\crcr}
   \writelabel{#1}\medskip \noindent\hskip -\spaceskip \ignorespaces }
\def\rwl@begin#1\cr{\rw@toks={#1\crcr}\rel@x
   \immediate\write\labelswrite{\the\rw@toks}\futurelet\n@xt\rwl@next}
\def\rwl@next{\ifx\n@xt\rwl@end \let\n@xt=\rel@x
      \else \let\n@xt=\rwl@begin \fi \n@xt}
\let\rwl@end=\rel@x
\def\writelabel#1{\immediate\write\labelswrite{\noexpand\labelbegin}
     \rwl@begin #1\cr\rwl@end
     \immediate\write\labelswrite{\noexpand\labelend}}
\newtoks\FromAddress         \FromAddress={}
\newtoks\sendername          \sendername={}
\newbox\FromLabelBox
\newdimen\labelwidth          \labelwidth=6in
\def\makelabels{\afterassignment\Makelabels \sendersname=}
\def\Makelabels{\FRONTPAGE \letterinfo={\hfil } \MakeFromBox
     \immediate\closeout\labelswrite  \input \jobname.lab\vfil\eject}
\let\labelend=\rel@x
\def\labelbegin#1\labelend{\setbox0=\vbox{\ialign{##\hfil\cr #1\crcr}}
     \MakeALabel }
\def\MakeFromBox{\gl@bal\setbox\FromLabelBox=\vbox{\Tenpoint
     \ialign{##\hfil\cr \the\sendername \the\FromAddress \crcr }}}
\def\MakeALabel{\vskip 1pt \hbox{\vrule \vbox{
        \hsize=\labelwidth \hrule\bigskip
        \leftline{\hskip 1\parindent \copy\FromLabelBox}\bigskip
        \centerline{\hfil \box0 } \bigskip \hrule
        }\vrule } \vskip 1pt plus 1fil }
\def\signed#1{\par \nobreak \bigskip \dt@pfalse \begingroup
  \everycr={\noalign{\nobreak
            \ifdt@p\vskip\signatureskip\gl@bal\dt@pfalse\fi }}%
  \tabskip=0.5\hsize plus \z@ minus 0.5\hsize
  \halign to\hsize {\strut ##\hfil\tabskip=\z@ plus 1fil minus \z@\crcr
          \noalign{\gl@bal\dt@ptrue}#1\crcr }%
  \endgroup \bigskip }
\newbox\letterb@x
\def\lettertext{\par \vskip\parskip \unvcopy\letterb@x \par }
\def\multiletter{\setbox\letterb@x=\vbox\bgroup
      \everypar{\vrule height 1\baselineskip depth 0pt width 0pt }
      \singlespace \topskip=\baselineskip }
\def\letterend{\par\egroup}
%
%
%
\newskip\frontpageskip
\newtoks\Pubnum   
\newtoks\Pubtype  \let\pubtype=\Pubtype
\newif\ifp@bblock  \p@bblocktrue
\def\PH@SR@V{\doubl@true \baselineskip=24.1pt plus 0.2pt minus 0.1pt
             \parskip= 3pt plus 2pt minus 1pt }
\def\PHYSREV{\papers\PhysRevtrue\PH@SR@V}

\def\titlepage{\FRONTPAGE\papers\ifPhysRev\PH@SR@V\fi
   \ifp@bblock\p@bblock \else\hrule height\z@ \rel@x \fi }
\def\nopubblock{\p@bblockfalse}
\def\endpage{\vfil\break}
\frontpageskip=12pt plus .5fil minus 2pt
\Pubtype={}
\Pubnum={}
\def\p@bblock{\begingroup \tabskip=\hsize minus \hsize
   \baselineskip=1.5\ht\strutbox \topspace-2\baselineskip
   \halign to\hsize{\strut ##\hfil\tabskip=0pt\crcr
       \the\Pubnum\crcr\the\date\crcr\the\pubtype\crcr}\endgroup}
\def\title#1{\vskip\frontpageskip \titlestyle{#1} \vskip\headskip }
\def\author#1{\vskip\frontpageskip\titlestyle{\twelvecp #1}\nobreak}
\def\andauthor{\vskip\frontpageskip\centerline{and}\author}

\def\address#1{\par\kern 5pt\titlestyle{\twelvepoint\it #1}}
\def\andaddress{\par\kern 5pt \centerline{\sl and} \address}

\def\abstract{\par\dimen@=\prevdepth \hrule height\z@ \prevdepth=\dimen@
   \vskip\frontpageskip\centerline{\fourteenrm ABSTRACT}\vskip\headskip }

%
%
%

\def\\{\rel@x \ifmmode \backslash \else {\tt\char`\\}\fi }
\def\sequentialequations{\rel@x \if\equanumber<0 \else
  \gl@bal\equanumber=-\equanumber \gl@bal\advance\equanumber by -1 \fi }
\def\journal#1&#2(#3){\begingroup \let\journal=\dummyj@urnal
    \unskip, \sl #1\unskip~\bf\ignorespaces #2\rm
    (\afterassignment\j@ur \count255=#3), \endgroup\ignorespaces }
\def\j@ur{\ifnum\count255<100 \advance\count255 by 1900 \fi
          \number\count255 }
\def\dummyj@urnal{%
    \toks@={Reference foul up: nested \journal macros}%
    \errhelp={Your forgot & or ( ) after the last \journal}%
    \errmessage{\the\toks@ }}

\def\topspace{\hrule height 0pt depth 0pt \vskip}

\def\Buildrel#1\under#2{\mathrel{\mathop{#2}\limits_{#1}}}
\def\becomes#1{\mathchoice{\becomes@\scriptstyle{#1}}
   {\becomes@\scriptstyle{#1}} {\becomes@\scriptscriptstyle{#1}}
   {\becomes@\scriptscriptstyle{#1}}}
\def\becomes@#1#2{\mathrel{\setbox0=\hbox{$\m@th #1{\,#2\,}$}%
        \mathop{\hbox to \wd0 {\rightarrowfill}}\limits_{#2}}}

\def\braket#1#2{\VEV{#1 | #2}}
\def\VEV#1{\left\langle #1\right\rangle}

\let\int=\intop         
\def\lsim{\mathrel{\mathpalette\@versim<}}
\def\gsim{\mathrel{\mathpalette\@versim>}}
\def\@versim#1#2{\vcenter{\offinterlineskip
        \ialign{$\m@th#1\hfil##\hfil$\crcr#2\crcr\sim\crcr } }}
\def\big#1{{\hbox{$\left#1\vbox to 0.85\b@gheight{}\right.\n@space$}}}
\def\Big#1{{\hbox{$\left#1\vbox to 1.15\b@gheight{}\right.\n@space$}}}
\def\bigg#1{{\hbox{$\left#1\vbox to 1.45\b@gheight{}\right.\n@space$}}}
\def\Bigg#1{{\hbox{$\left#1\vbox to 1.75\b@gheight{}\right.\n@space$}}}
\def\){\mskip 2mu\nobreak }
%
%
%
\let\sec@nt=\sec
\def\sec{\rel@x\ifmmode\let\n@xt=\sec@nt\else\let\n@xt\section\fi\n@xt}
\def\obsolete#1{\message{Macro \string #1 is obsolete.}}
\def\firstsec#1{\obsolete\firstsec \section{#1}}
\def\firstsubsec#1{\obsolete\firstsubsec \subsection{#1}}
\def\thispage#1{\obsolete\thispage \gl@bal\pagenumber=#1\frontpagefalse}
\def\thischapter#1{\obsolete\thischapter \gl@bal\chapternumber=#1}
\def\splitout{\obsolete\splitout\rel@x}
\def\prop{\obsolete\prop \propto }
\def\nextequation#1{\obsolete\nextequation \gl@bal\equanumber=#1
   \ifnum\the\equanumber>0 \gl@bal\advance\equanumber by 1 \fi}
\def\BOXITEM{\afterassigment\B@XITEM\setbox0=}
\def\B@XITEM{\par\hangindent\wd0 \noindent\box0 }
%
%
%
\def\phyzzx{PHY\setbox0=\hbox{Z}\copy0 \kern-0.5\wd0 \box0 X}
        
\everyjob{
        \input myphyx.tex }
\message{ by V.K.}
%
\catcode`\@=12 
%
%
%
%
%

\font\fourteenrm  =cmr10 scaled\magstep2
\font\twelverm    =cmr12
\font\ninerm      =cmr9
\font\sixrm       =cmr6

\font\fourteenbf  =cmbx10 scaled\magstep2
\font\twelvebf    =cmbx12
\font\ninebf      =cmbx9
\font\sixbf       =cmbx6
\font\seventeeni  =cmmi10 scaled\magstep3    \skewchar\seventeeni='177
\font\fourteeni   =cmmi10 scaled\magstep2     \skewchar\fourteeni='177
\font\twelvei     =cmmi12                       \skewchar\twelvei='177
\font\ninei       =cmmi9                          \skewchar\ninei='177
\font\sixi        =cmmi6                           \skewchar\sixi='177
\font\seventeensy =cmsy10 scaled\magstep3    \skewchar\seventeensy='60
\font\fourteensy  =cmsy10 scaled\magstep2     \skewchar\fourteensy='60
\font\twelvesy    =cmsy10 scaled\magstep1       \skewchar\twelvesy='60
\font\ninesy      =cmsy9                          \skewchar\ninesy='60
\font\sixsy       =cmsy6                           \skewchar\sixsy='60

\font\fourteenex  =cmex10 scaled\magstep2
\font\twelveex    =cmex10 scaled\magstep1

\font\fourteensl  =cmsl10 scaled\magstep2
\font\twelvesl    =cmsl12
\font\ninesl      =cmsl9

\font\fourteenit  =cmti10 scaled\magstep2
\font\twelveit    =cmti12
\font\nineit      =cmti9
\font\fourteentt  =cmtt10 scaled\magstep2
\font\twelvett    =cmtt12
\font\fourteencp  =cmcsc10 scaled\magstep2
\font\twelvecp    =cmcsc10 scaled\magstep1
\font\tencp       =cmcsc10
%
%
\def\fourteenf@nts{\relax
    \textfont0=\fourteenrm          \scriptfont0=\tenrm
      \scriptscriptfont0=\sevenrm
    \textfont1=\fourteeni           \scriptfont1=\teni
      \scriptscriptfont1=\seveni
    \textfont2=\fourteensy          \scriptfont2=\tensy
      \scriptscriptfont2=\sevensy
    \textfont3=\fourteenex          \scriptfont3=\twelveex
      \scriptscriptfont3=\tenex
    \textfont\itfam=\fourteenit     \scriptfont\itfam=\tenit
    \textfont\slfam=\fourteensl     \scriptfont\slfam=\tensl
    \textfont\bffam=\fourteenbf     \scriptfont\bffam=\tenbf
      \scriptscriptfont\bffam=\sevenbf
    \textfont\ttfam=\fourteentt
    \textfont\cpfam=\fourteencp }
\def\twelvef@nts{\relax
    \textfont0=\twelverm          \scriptfont0=\ninerm
      \scriptscriptfont0=\sixrm
    \textfont1=\twelvei           \scriptfont1=\ninei
      \scriptscriptfont1=\sixi
    \textfont2=\twelvesy           \scriptfont2=\ninesy
      \scriptscriptfont2=\sixsy
    \textfont3=\twelveex          \scriptfont3=\tenex
      \scriptscriptfont3=\tenex
    \textfont\itfam=\twelveit     \scriptfont\itfam=\nineit
    \textfont\slfam=\twelvesl     \scriptfont\slfam=\ninesl
    \textfont\bffam=\twelvebf     \scriptfont\bffam=\ninebf
      \scriptscriptfont\bffam=\sixbf
    \textfont\ttfam=\twelvett
    \textfont\cpfam=\twelvecp }
\def\tenf@nts{\relax
    \textfont0=\tenrm          \scriptfont0=\sevenrm
      \scriptscriptfont0=\fiverm
    \textfont1=\teni           \scriptfont1=\seveni
      \scriptscriptfont1=\fivei
    \textfont2=\tensy          \scriptfont2=\sevensy
      \scriptscriptfont2=\fivesy
    \textfont3=\tenex          \scriptfont3=\tenex
      \scriptscriptfont3=\tenex
    \textfont\itfam=\tenit     \scriptfont\itfam=\seveni  
    \textfont\slfam=\tensl     \scriptfont\slfam=\sevenrm 
    \textfont\bffam=\tenbf     \scriptfont\bffam=\sevenbf
      \scriptscriptfont\bffam=\fivebf
    \textfont\ttfam=\tentt
    \textfont\cpfam=\tencp }
%
%
\PHYSREV
\def\fine{\endpage\refout}
\def\pfi{{\cal P}_\Phi(I)}
\def\dquadro{{d^2\over dx^2}}
\def\cald{{\cal D}}
\def\calj{{\cal J}}
\def\caln{{\cal N}}
\def\calt{{\cal T}}
\def\sx{\sigma^x}
\def\sy{\sigma^y}
\def\sz{\sigma^z}
\def\sp{s^+}
\def\sm{s^-}
\def\Ima{{\rm Im}\,}
\def\as{\alpha_s}
\def\bs{\beta_s}
\def\ave#1{\left\langle{#1}\right\rangle}
\def\dkls{{ \left(I_0\,\sin\,2\pi\Phi\right) }}
\def\eal{{ e^{-{ L\over 2\ell}} }}

\titlepage
\title{Statistics of the One-Electron Current in a One-Dimensional Mesoscopic
Ring at Arbitrary Magnetic Fields}
\author{Alberto S. Cattaneo\foot{Also Sezione INFN di Milano,
20133 Milano, Italy.
E.mail: {\it cattaneo@vaxmi.mi.infn.it}.}}
\address{Dipartimento di Fisica, Universit\`a degli Studi di Milano,
20133 Milano, Italy}
\author{Andrea Gamba\foot{Also Sezione INFN di Pavia,
27100 Pavia, Italy.
E.mail: {\it gamba@polito.it}.}}
\address{Dipartimento di Matematica, Politecnico di Torino, 10129 Torino,
Italy}
\andauthor{Igor V. Kolokolov\foot{Permanent address: Budker Institute of
Nuclear Physics, 630090 Novosibirsk, Russia.
E.mail: {\it kolokolov@vaxmi.mi.infn.it}}}
\address{Sezione INFN di Milano, 20133 Milano,
Italy}
\abstract{
The set of moments and the distribution function of the one-electron current
in a one-di\-men\-sio\-nal disordered ring with arbitrary magnetic flux
are calculated.}
\endpage\pagenumber=1
\chapter{Introduction}
\REF\BIL{M. B\"uttiker, Y. Imry and R. Landauer\journal Phys. Lett.
&A96(83)365.}
\REF\LDDB{L.P. Levy, G. Dolan, J. Dunsmuir and H. Bouchiat\journal Phys.
Rev. Lett. &64(90)2074.}
\REF\CWBKGK{V. Chandrasekhar, R.A. Webb, M.J. Brady, M.B. Ketchen, W.J.
Gallagher and A. Kleinsasser\journal Phys. Rev. Lett. &67(91)3578.}
\REF\Sch{A. Schmid\journal Phys. Rev. Lett. &66(91)80.}
\REF\OR{F. von Oppen and E.K. Riedel\journal Phys. Rev. Lett.
&66(91)84.}
\REF\AGI{B.L. Altschuler, Y. Gelfen and Y. Imry\journal Phys. Rev. Lett.
&66(91)88.}
\REF\ES{V. Eckern and A. Schmid\journal Ann. der Physik &2(93)180.}
\REF\MWL{A. M\"uller-Groeling, H.A. Weidenm\"uller and C.H.
Lewenkopf\journal Eur. Lett. &22(93)193.}
\REF\Kol{I.V. Kolokolov
\journal Journ. of Exp. and Theor. Phys. (formerly Sov. Phys. JETP)
&76(93)6.}
\REF\Dor{O. Dorokhov\journal Sov. Phys. JETP &74(92)518.}
\REF\AIMW{A. Altland, S. Iida, A. M\"uller-Groeling and H.A.
Weidenm\"uller\journal Annals of Phys. (NY) &219(92)148.}
\REF\AR{A.A. Abrikosov and I.A.
Ryzhkin\journal Adv. in Phys. &27(78)146.}\

\par\noindent
The theoretical work of Ref.~\BIL,
together with recent experimental results [\LDDB, \CWBKGK]
have
stirred a strong interest toward the problem of the
persistent current in a mesoscopic metal ring immersed in a magnetic field.
This current has been computed using various approaches
[\Sch-\AIMW].
In some works
[\Sch-\ES]
it was considered the experimental
situation of a ring of finite thickness,
such that the number of transverse
channels be much greater than one. This allowed the authors
to use the methods of
weak-localization theory.
In this case one has also to take into
account the electron-electron interaction [\MWL].
\par
On the other hand, the computation of the one-electron current in an
idealized one-dimensional disordered ring is also of interest, at least
from the theoretical point of view. One-dimensional localization
effects lead in this case to a non-trivial current dependence on the
magnetic flux (see Ref. \Kol\ and below).
In Ref.~\Dor\ such a calculation was performed, but only in the
weak magnetic field case.
In Ref. \AIMW\ the
one-electron current averaged over an ensemble of rings was
derived non-perturbatively using Grassmann matrix
integration. However, the unexplicit form of the resulting expression and
the need of
tedious computations do not allow to check the validity of
some approximations.
\par
In Ref. \Kol\ it was developed a new path integral approach to the
study of one-dimensional
localization. Along with the multipoint density
correlators, this new method allowed to compute the averaged absolute values
$\ave{|I|}$ of the
one-electron current $I$ in a disordered metal ring with arbitrary
magnetic flux.
\par
In the present paper we show that the method
introduced in Ref.~\Kol\ allows us to reconstruct
completely the distribution function $\pfi$ over
an ensemble of one-dimensional rings with given magnetic flux
$\Phi$.
Simple explicit expressions for the moments
$\ave{I^{2n}}$ of the current $I$ are also obtained

\chapter{Path Integral Representation}
\noindent
Let us recall the main steps of the path-integral approach
introduced in Ref.~\Kol.
The Schr\"odinger equation
$$
(\hat H -k^2)\psi = \left({-\dquadro + U(x) - k^2}\right)\psi = 0
\eqn\uno
$$
maps the two-dimensional space of the initial conditions
$(\psi'(x_0)+ik\psi(x_0)$, $\psi'(x_0)-ik\psi(x_0))$ to the
two-dimensional space of the solutions at the point $x$ through the matrix
$$
T(x,x_0) = e^{ik(x-x_0)\sz} \calt(x,x_0) e^{ik(x-x_0)\sz},
\eqn\due
$$
where $\calt(x,x_0)$ obeys the following first-order equation:
$$
{d\over dx}\calt = (i\varphi(x)s^z + \zeta^+(x)\sm + \zeta^-(x)\sp ) \calt,
\eqn\tre
$$
and
$$
\varphi(x) = -{1\over k} U(x),\qquad
\zeta^\pm(x) = \pm {i\over 2k} U(x) e^{\pm 2ikx}.
\eqn\quattro
$$
Here $s^z=\sz/2$ and $s^\pm=(\sx\pm i\sy)/2$ are the usual
spin operators.
It has been shown [\AR]
(see also Ref. \Kol\ for greater detail) that in the limit
$$
k\ell\gg 1,
\eqn\cinque
$$
where $\ell$ is the mean free path, the fields $\varphi(x)$ and $\zeta^\pm(x)$
can be considered as statistically independent. If the initial
potential $U(x)$ is a Gaussian random function of $x$ with
correlator
$$
\ave{U(x)U(x')} = D\delta(x-x'),
\eqn\sei
$$
then the averaging weight over the fields $\varphi(x)$ and $\zeta^\pm(x)$ has
the form:
$$
\cald\varphi(x)\ \cald\zeta^\pm(x)\ \exp \left({
-\ell\cdot\int dx\,[{1\over 8}\varphi^2(x)+\zeta^+(x)\zeta^-(x)]}\right),
\eqn\sette
$$
where $\ell=4k^2/D$ is the localization length.
\REF\Koluno{I.V. Kolokolov\journal Phys. Lett. &A114(86)99.}
\REF\KP{I.V. Kolokolov and E.V. Podivilov\journal Sov. Phys. JETP
&68(89)119.}
\REF\Koldue{I.V. Kolokolov\journal Ann. of Phys. (NY) &202(90)165.}
\REF\akerman{E.~Akkermans, A.~Auerbach, J.E.~Avron and B.~Shapiro
\journal Phys. Rev. Lett. &66(91)76.}
It has been shown [\Kol,\Koluno-\Koldue]
that the following change of variables in \sette
$$
\eqalign{
i\varphi &= i\rho +2\psi^+\psi^-,\cr
\zeta^- &= \dot \psi^- -i\rho\psi^- - \psi^+(\psi^-)^2,\cr
\zeta^+ &= \psi^+\cr}
\eqn\otto
$$
brings the operator $\calt(x,x_0)$
in the form of a product of usual matrix exponentials:
$$
\calt(x,x_0) = \exp\left[{\sp\psi^-(x)}\right]
\exp\left[{is^z\int_{x_0}^xdt\,\rho(t)}\right]
\exp\left[{\sm\int_{x_0}^xdt\,\psi^+(t)
e^{i\int_{x_0}^tdt'\,\rho(t')}}\right].
\eqn\nove
$$
This statement can be checked
by deriving the evolution equation for the operator \nove\ and comparing
it with \tre.
The field $\psi^-(x)$ is assumed to obey the initial condition
$
\psi^-(x_0) = 0,
$
thus providing the equality $\calt(x_0,x_0)=1$.
Under a proper regularization, which is required by physical
considerations (see Ref.~\Kol), the Jacobian of the transformation \otto\ is
seen to be equal to
$$
\calj\propto \exp\left({-{i\over2}\int_{-L}^L dt\,\rho(t)}\right).
\eqn\dieci
$$
The surface of integration in the space of the complex fields
$\varphi,\zeta^\pm$ is defined by the equations
$$
\Ima\varphi = 0,\qquad\zeta^-=(\zeta^+)^\ast,
\eqn\undici
$$
where $*$ denotes complex conjugation. This surface can be deformed
to the standard one
$$
\Ima\rho = 0,\qquad\psi^- = (\psi^+)^\ast,
\eqn\dodici
$$
if all the quantities to be averaged are written in a form which allows
analytical continuation from the surface \undici\ to the whole complex
space of field configurations (for more details see Ref. \Kol\ and
\Koluno). This requirement turns out to be fairly constructive.
\chapter{Calculation of the Current Moments and of the Current Distribution
Function}
\noindent
In an appropriate gauge the wave-function of an electron moving
in a metal ring of size $2L$ immersed in a
magnetic flux $\Phi$, measured in units of flux quanta,
obeys eqn. \uno. Topology and flux dependence
are then encoded in the boundary conditions
$$
(\psi'(L)\pm ik\psi(L)) = e^{2\pi i\Phi} (\psi'(-L)\pm ik\psi(-L)).
\eqn\tredici
$$
The mean value of a function $f(I)$ of the current $I$ can be defined as
follows:
$$
\ave{f(I)}=
\ave{{2\pi k\over L}\sum_n\delta(E-E_n)f(j_n)},\qquad
\hbox{ where }
\qquad j_n = -{1\over 2\pi}{\partial E_n\over\partial\Phi}.
\eqn\quattordici
$$
Here $E=k^2$ is the electron energy and $E_n$ are the eigenvalues of the
Hamiltonian \uno\ with boundary conditions \tredici, which
can be written in terms of the matrix $T\equiv T(L,-L)$:
$$
\det(T-e^{{2\pi i\Phi}}) = 0.
\eqn\sedici
$$
The matrix $\calt\equiv\calt(L,-L)$ satisfies the ``unitarity" conditions:
$$
\matrix{
\sz\calt^{\dag}\sz =\calt^{-1},\hfill
&\hfill \det\calt = 1,\cr}
\eqn\diciassette
$$
and therefore admits
the following parametrization:
$$
\calt=\left(\matrix{
e^{ i\as}\cosh\Gamma & e^{ i\bs}\sinh\Gamma\cr
e^{-i\bs}\sinh\Gamma & e^{-i\as}\cosh\Gamma\cr}\right).
\eqn\diciassettebis
$$
Here $\as,\bs$ and $\Gamma$ are by construction (see Ref. \CWBKGK) slowly
varying real functions of $L$. Substituting the parametrization
\diciassettebis\ into \due, we obtain from \sedici\ the equation
determining the set of the eigenvalues $E_n$ [\Dor]:
$$
\tau(E)\equiv\cosh\Gamma\,\cos(\as+kL)=\cos{2\pi\Phi}.
\eqn\diciotto
$$
Let us start the computation of $\ave{I^{2n}}$:
$$
\ave{I^{2n}}
= \ave{ {2\pi k\over L}\sum_n\delta(E-E_n)j_n^{2m}}
= \ave{
{2\pi k\over L}\delta(\tau(E)-\cos{2\pi\Phi})
{\sin^{2m}{2\pi\Phi}\over |\tau'(E)|^{2m-1}}
}.
\eqn\diciannove
$$
Here \sedici\ and \diciotto\ have been taken into account. The
$\delta$-function can be eliminated in \diciannove\ using the following
consideration: for $kL\gg 1$ the result of the average \diciannove\ does
not change when $L$ varies on a scale much less than $\ell$. Then
$\ave{I^{2m}}$ must coincide with its average over an
interval $\Delta L$ of lengths $L$:
$$
\ave{I^{2m}}_L = {1\over\Delta L} \int_L^{L+\Delta L}
dL\,\ave{I^{2m}}_L,
\qquad\hbox{ where }\qquad
{1\over k}\ll\Delta L \ll l.
\eqn\venti
$$
We can interchange the order of the two averages; then, using the (approximate)
constancy of the variables $\Gamma$, $\as$ and $\bs$ on the interval
$\Delta L$, we obtain\rlap:\foot{Such a procedure,
which was proposed in Ref. \Kol,
seems to be equivalent to the rings-ensemble averaging of Ref. \AIMW.
See also Ref.~\akerman.}
$$
\ave{I^{2m}} = \dkls^{2m}\ave{1\over(\sinh^2\Gamma+\sin^2{2\pi\Phi})^m},
\eqn\ventidue
$$
where we have set $I_0=2k/L$.
Eqn. \ventidue\ can be rewritten in the form:
$$
\ave{I^{2m}} = {2\over(m-1)!}\dkls^{2m}\int_0^\infty d\mu\,\mu^{2m-1}
\ave{e^{-\mu^2(\sinh^2\Gamma+\sin^2{2\pi\Phi})}}.
\eqn\ventitre
$$
It is important to notice  that $\sinh^2\Gamma$ can be expressed in terms of
the elements of the matrix $\calt$ without using any complex conjugation:
$$
\sinh^2\Gamma=(\matrix{1 & 0\cr})\calt^t\sm\calt
\left(\matrix{1\cr 0\cr}\right),
\eqn\ventiquattro
$$
where $t$ denotes the usual matrix transposition. Thus
the above mentioned
analytic
continuation from the surface \undici\ is possible. Substituting \nove\
into \ventitre\ we obtain:
$$
\sinh^2\Gamma=\psi^-(L)\int_{-L}^Ldt\,\psi^+(t)e^{-i\int_t^Ldt'\,\rho(t')}.
\eqn\venticinque
$$
The average \ventitre\ is performed using the weight
$$
\cald\rho\ \cald\psi^\pm\ e^{-S'(\rho,\psi^\pm)},
\eqn\ventisei
$$
where the action $S'(\rho,\psi^\pm)$ is obtained from \sette\ after the
substitution \otto, taking into account the Jacobian \dieci, and reads
$$
S'(\rho,\psi^\pm)=\ell\,\int_{-L}^Ldx\,\left[{
{1\over 8}\rho^2+\psi^+\dot\psi^- -{3\over 2}i\rho\psi^+\psi^-
-{3\over 2}(\psi^+\psi^-)^2}\right]
+{i\over2}\int_{-L}^Ldx\,\rho.
\eqn\ventisette
$$
\REF\PW{A.M. Polyakov and P.B.
Wiegmann\journal Phys. Lett. &B131(83)121.}
\noindent
This action has the form of a $(0+1)$--dimensional Schwinger model and the
$\cald\psi^\pm$--integration in \ventitre\ can be performed using the
so-called ``bosonization" method [\PW],
representing
$$
\exp\left[
{3\over 2}\ell\,\int_{-L}^Ldx\,(\psi^+\psi^-)^2 \right]=
\int\cald\eta\ \exp\left[{-{3\over 2}\ell\,\int_{-L}^Ldx\,
(\eta^2+2\eta\psi^+\psi^-)}\right]
\eqn\ventotto
$$
and:
$$
e^{-\mu^2\sinh^2\Gamma}=
{1\over\pi}\int dz\,dz^\ast e^{-|z|^2} \exp\left({-i\mu z\psi^-(L)
-i\mu
z^\ast\int_{-L}^Ldx\,\psi^+(x)e^{-i\int_x^Ldt\,\rho(t)}
}\right).
\eqn\ventinove
$$
We eliminate the $\rho\,\psi^+\psi^-$ and $\eta\,\psi^+\psi^-$
interaction terms through the following gauge transformation:
$$
\psi^\pm(x)\longrightarrow\psi^\pm(x)
\exp\left[{\pm{3\over 2}\int_{-L}^xdt\,(2\eta-i\rho)}\right],
\eqn\trenta
$$
which has the Jacobian
$$
\calj_R\propto
\exp\left[{-{3\over 4}\int_{-L}^Ldt\,(2\eta-i\rho)}\right].
\eqn\trentuno
$$
The $\cald\psi^\pm$--integration becomes Gaussian and can be easily
performed. Introducing the variable $\xi(x)$ and denoting the
$x$-derivative with a dot we have
$$
\dot\xi = -3\eta+{i\over 2}\rho,\qquad\quad
\xi(L)=0,\qquad\quad
\cald\rho\ \cald\eta\propto\cald\rho\ \cald\xi,
\eqn\trentadue
$$
and performing the Gaussian $\cald\rho$-integration, we come to the
following expression for $\ave{I^{2m}}$:
$$
\eqalign{
\ave{I^{2m}}
&={2\over\pi(m-1)!}\dkls^{2m}\int_0^\infty d\mu\,
\mu^{2m-1}\int dz\,dz^\ast\, e^{-\mu^2\sin^2{2\pi\Phi}-|z|^2}\cdot\cr
&\cdot\caln e^{-{L\over 2\ell}}
\int_{\xi(L)=0}\cald\xi\
\exp\left[{
-{\ell\over 4}\int_{-L}^Ldx\,(\dot\xi^2+
{4\over\ell^2}\mu^2|z|^2e^{-\xi})}\right]
e^{-{\xi(-L)\over2}}=\cr
&={\ell^{2m}\over 2^{2m-1}(m-1)!}\dkls^{2m} e^{-{L\over 2\ell}}
\int_0^\infty {dr\over r^{2m-1}} e^{-r^2}
\ave{\Upsilon_2^{(m)}(\xi,r) \big|e^{-2L\hat H}\big| \Upsilon_1(\xi)}\cr}
\eqn\trentatre
$$
in terms of usual quantum-mechanical matrix elements with Hamiltonian
$$
\hat H = -{1\over\ell}{d^2\over d\xi^2}-{\ell\over 4}e^{-\xi},
\eqn\trentaquattro
$$
where the ket- and bra- wave funtions are:
$$
\matrix{
\Upsilon(\xi) = e^{-{\xi\over2}},\hfill
&\hfill\Upsilon_2^{(m)}(\xi,r) = e^{-\xi(m-1/2)}
\exp\left({-{\ell^2\sin^2{2\pi\Phi} e^{-\xi}\over 4 r^2}}\right).\cr}
\eqn\trentacinque
$$
The factor $\caln$ in
\trentatre\ is the normalization of the standard Feynman--Kac
path-integral; together with $\eal$ it provides the equality $\ave1=1$.
Using the complete set of eigenfunctions of $\hat H$
$$
\matrix{
f_\nu(\xi)={2\over\pi}\sqrt{\nu\sinh{2\pi\nu}}\;K_{2i\nu}\left({\ell\,
e^{-{\xi\over2}}}\right),\hfill
&\hat Hf_\nu(\xi)=-{1\over\ell}\,\nu^2f_\nu(\xi),
&\hfill\braket{f_\nu}{f_{\nu'}}=\delta(\nu-\nu'),\cr}
\eqn\trentasei
$$
where $K_\mu$ is the standard notation for the modified Bessel function,
we obtain, after some arithmetic:
$$
\ave{I^{2m}}
=\dkls^{2m}{(-1)^{m-1}\over (m-1)!}\left({\partial\over\partial
(\sin^2{2\pi\Phi})}\right)^{m-1}\ave{I^2\dkls^{-2}},
\eqn\trentasette
$$
and
$$
\eqalign{
\ave{I^2}
&=
{
2\eal
\over
\sqrt{\pi}(2L/\ell)^{3/2}
}
\dkls^2\int_0^\infty dx\,
{
x\,e^{-{\ell\over 2L}x^2}
\over
\sqrt{\sinh^2x+\sin^2\,2\pi\Phi}
}
\cdot
\cr
&\cdot\log
\left[
1+2
{\sinh^2x\over\sin^2\,2\pi\Phi}
+2
\left(
{\sinh^2x\over\sin^2\,2\pi\Phi}
+
{\sinh^4x\over\sin^4\,2\pi\Phi}
\right)^{1/2}\right].\cr}
\eqn\trentotto
$$
In order to reconstruct the distribution function $\pfi$ we use the following
identity:
$$
{\cal P}_\Phi(J)={1\over\pi}\lim_{\epsilon\to0^+}\Ima\int
{\pfi dI\over J-I+i\epsilon}={1\over\pi J}\lim_{\epsilon\to0^+}\Ima
\sum_{n=0}^{\infty}{\ave{I^{2m}}\over(J+i\epsilon)^{2n}}.
\eqn\trentanove
$$
It is convenient to compute the sum in \trentanove\ considering it as the
result of an analytical continuation in $J$ from the imaginary positive
semiaxis. Substituting \trentasette\ into \trentanove\ we see that for
such $J$ the summation gives a well-defined translation operator in the
variable $\sin^2{2\pi\Phi}$ acting on $\ave{I^2\dkls^{-2}}$. Performing the
analytical continuation to the real axis we obtain
$$
\pfi = 0,\qquad\hbox{for }|I|>I_0
\eqn\quaranta
$$
and
$$
\pfi={2\eal\over\sqrt\pi({2 L/\ell})^{3/2}} {(I_0\,\sin\,2\pi\Phi)^2\over|I|^3}
\int_{\lambda(I)}^{+\infty} dx\, {xe^{-{\ell\over 2L}{x^2}}
\over\sqrt{\sinh^2x-\sinh^2{\lambda(I)}}},
\qquad\hbox{for }|I|<I_0,
\eqn\quarantuno
$$
where
$$
I_0 \equiv{2k\over L}\qquad
\hbox{ and }
\qquad
\lambda(I)={\rm sinh}^{-1}\,\left[{
|\sin{2\pi\Phi}|\left({{I_0^2\over I^2}-1}\right)^{1/2}}\right].
\eqn\quarantunobis
$$
It can be checked that this distribution function reproduces all the
moments $\ave{I^{2m}}$ as well as the result for $\ave{|I|}$ obtained in
Ref.~\Kol.

In the limit $I\rightarrow 0$
we get
$$
\lambda\sim\log\left(2{|I_0\sin{2\pi\Phi}|\over|I|}\right)
\eqn\nuovauno
$$
and
$$
\pfi\sim
{\lambda \exp\left({-{\ell\over{2 L}}(\lambda-{{ L}\over \ell})^2}\right)
\over
|I|\,({2 L/\ell})^{3/2}}\,
{\Gamma\left({\lambda\over{2 L/\ell}}+{1\over 2}\right)
\over
\Gamma\left({\lambda\over{2 L/\ell}}+1\right)}
\eqn\nuovadue
$$
in the proximity of the maximum $\lambda\sim{{ L}\over \ell}$.
In the limit ${ L/\ell}\rightarrow +\infty$
we get for the quantity $\lambda$
the normal distribution
$$
\pfi\,dI=
{e^{-{\ell\over 2 L}(\lambda-{L\over \ell})^2}
\over\sqrt{2\pi{L/\ell}}}
\,d\lambda.
\eqn\nuovatre
$$
Thus in the thermodynamic limit the fluctuations of $\lambda$ are
suppressed and $\lambda$ becomes a non-random quantity.
\REF\melnikov{V.I.~Mel'nikov
\journal Sov. Phys. Solid State
&23(81)444.}
This fact is in a deep connection with an earlier result~[\melnikov]
about the asymptotically normal distribution of the logarithm of the
static resistivity (see also Ref.~\AR).
In both cases we are dealing with the response of the system to an
external field.
In our case $\sin^2{2\pi\Phi}$ can formally assume an arbitrary value
and in some sense we are considering a non-linear response.
However we see from~\nuovatre\ that in the limit ${L/\ell}\rightarrow +\infty$
the response becomes effectively linear.

When $I\rightarrow I_0$~\quarantuno\ gives
$$
{\cal P}_\Phi(I_0)={2\eal\over\sqrt\pi({2 L/\ell})^{3/2}}
{\sin^2{2\pi\Phi}\over I_0}\int_0^\infty dx\,{xe^{-{\ell\over 2L}{x^2}}
\over\sinh x}.
$$
It is worth noting that in the limit
${2\pi\Phi}\rightarrow 0$,
when $\ell/L<\infty$ is fixed,
all the moments of the current $I$ tend to zero.
This seems natural since in a given potential without symmetries
($\ell/L<\infty$) and with zero magnetic field all the stationary
states of the electron in the ring can be described by real
wave functions.
The corresponding quantum-mechanical expectation values of the current
operator are equal to zero.
On the other hand, if we take simultaneously the limit
$\ell/L\rightarrow\infty$
(free motion case) we can obtain a non-zero result.

Let us also note that the formal substitution
$I_0=1$,
$\sin^2{2\pi\Phi}=1$
and
$I^2\rightarrow T$
in~\quarantuno\ gives us the distribution function for the
transmission coefficient $T$.
\REF\kirkman{P.D.~Kirkman and J.B.~Pendry
\journal J. Phys. C: Solid State Physics
&17(84)5707.}
\REF\pendry{J.B.~Pendry, A.~MacKinnon and P.J.~Robers
\journal Proc. Roy. Soc.
&437(92)6.}
In the limit ${L/\ell}\rightarrow\infty$
this reproduces the known results [\kirkman,\pendry]
for the moments $\ave{T^n}$,
but our formula is valid for finite values of $L/\ell$ as well
(the only limitation is that the sample length $2L$ and the localization
length $\ell$ be great in comparison to the wavelength $1/k$).

\bigskip
\centerline{\bf Acknowledgments}
\bigskip\noindent
We would like to thank M. Martellini for his interest in this work
and for his continuos support.
We are also grateful to the referee for his observations
which helped us to improve the text of this paper.

\fine\bye